\documentstyle[12pt]{article}
\def\rtimes{\mbox{$\times\!\rule{0.3pt}{1.1ex}\,$}} 
\def\o{{\sb{(1)}}} 
\def\t{{\sb{(2)}}} 
\def\th{{\sb{(3)}}} 
\def\f{{\sb{(4)}}} 
\def\bo{{}^{\bar{(1)}}}\def\bt{{}^{\bar{(2)}}} 
\def\id{{\rm id}} 
\def\ker{{\rm ker}} 
\def\im{{\rm Im}} 
 
\def\ad{{\rm Ad\sb{R}}} 
\def\S{{\rm S}} 
\def\CF{{\cal F}} 
\def\CA{{\cal A}} 
\def\CR{{\cal R}} 
 
\def\CG{{\cal G}} 
 
\def\C{{\bf C}} 
\def\Z{{\bf Z}} 
 
\def\deltens{{\Delta\sb R\sp\otimes}} 
\def\delad{{\Delta\sb R\sp{\rm ad}}} 
\def\tens{{\otimes}} 
\def\proof{\goodbreak\noindent{\bf Proof.\quad}} 
\def\endproof{{\ $\Box$}\bigskip } 
\def\dia{{\ $\Diamond$}} 
\def\eps{{\epsilon}} 
\def\tauo{{\tau\sp{(1)}}}\def\taut{{\tau\sp{(2)}}} 
\def\tauos{{\tau'\sp{(1)}}} 
\def\tauts{{\tau'\sp{(2)}}} 
 
\newtheorem{prop}{Proposition}[section] 
\newtheorem{lemma}[prop]{Lemma} 
\newtheorem{thm}[prop]{Theorem} 
\newtheorem{df}[prop]{Definition} 
\newtheorem{rk}[prop]{Remark} 
\newtheorem{cor}[prop]{Corollary} 
\newtheorem{ex}[prop]{Example}

\font\elevenbf=cmbx10 scaled\magstep 1 
\font\elevenrm=cmr10 scaled\magstep 1 
\font\elevenit=cmti10 scaled\magstep 1

\renewenvironment{thebibliography}[1] 
 { \elevenrm  
    \begin{list}{\arabic{enumi}.} 
     {\usecounter{enumi} \setlength{\parsep}{0pt} 
            \setlength{\itemsep}{3pt} \settowidth{\labelwidth}{#1.} 
            \sloppy 
          }}{\end{list}} 

\begin{document} 
\thispagestyle{empty} 
~
\vskip 1in 
{\begin{center} 
\baselineskip 22pt 
{\Large \bf Translation Map in Quantum Principal Bundles}   
\vskip 0.5in 
\baselineskip 13pt 
Tomasz Brzezi\'nski 
 \\[.3in] 
{\em  Physique Nucl\'eaire 
Th\'eorique et Physique Math\'ematique\\ Universit\'e Libre de Bruxelles\\  
Campus de la Plaine CP 229\\  B-1050 Brussels, 
Belgium\footnote{Current address: Department of Applied Mathematics
and Theoretical Physics, University of Cambridge, Cambridge CB3 9EW,
U.K. E-mail: T.Brzezinski@damtp.cam.ac.uk} 
} 
\vspace{24pt} 
 
February 1995 
\end{center} 
\vspace{1in} 
\baselineskip 14pt 
 
\begin{quote}ABSTRACT The notion of a translation map in a quantum 
principal bundle is introduced. A translation map is then used to 
prove that the cross sections of a quantum fibre bundle $E(B,V,A)$ 
associated to a quantum principal bundle $P(B,A)$ are in 
bijective correspondence with  equivariant maps $V\to P$, and that a 
quantum principal bundle is trivial if it admits a cross section which 
is an algebra map. 
The vertical automorphisms and gauge transformations  
of a quantum principal bundle are discussed. In particular it is shown 
that vertical automorphisms are in bijective correspondence with 
$\ad$-covariant maps $A\to P$. 
\end{quote} 
} 
\newpage 
\baselineskip 19pt

\section{Introduction} 
Quantum fibre bundles generalise the concept of a 
fibre bundle \cite{husemoller1} which plays a vital r\^ole both 
in mathematics and physics \cite{booss1,gockeler1}. The main idea of 
this generalisation is to replace the algebra of functions on a 
structure group by a Hopf algebra \cite{sweedler1} or a quantum group 
\cite{drinfeld1}, and the algebras of functions on the base manifold, 
fibre etc.~by non-commutative algebras. Such a generalisation of a 
fibre bundle was proposed  by S. Majid and the author in  
\cite{brzezinski6}. The general considerations of \cite{brzezinski6} 
were 
illustrated by reference to an example of a canonical connection in 
the quantised Hopf bundle, i.e.~by the deformation of a configuration 
that is known in physics as the Dirac monopole. The generalisation of 
fibre bundles proposed in \cite{brzezinski6} was then developed, 
partly independently, by various authors 
\cite{pflaum1,hajac1,durdevic1,budzynski1,chu1}.  Although our work in 
\cite{brzezinski6} was motivated by physics, i.e.~we aimed at a 
quantum group generalisation of gauge theories, the constructed 
objects seem to have more fundamental meaning reaching far beyond this 
particular application. It is 
therefore important not only to study physical constructions that may 
result from the quantum group generalisation of gauge theories but 
also to analyse the geometric structure of quantum fibre bundles.  
 
In this paper we show that quantum fibre bundles enjoy some properties 
similar to the properties  of classical fibre bundles. In 
particular we prove that cross sections of a quantum fibre bundle $E$ 
associated to a quantum principal bundle $P$ are in one-to-one 
correspondence with equivariant maps defined on the fibre of $E$ with 
values in $P$. We then deduce that if a quantum principal bundle has a 
cross section which is an algebra map, then the bundle is trivial. We 
also show how to interpret gauge transformations of a non-trivial 
quantum principal bundle in terms of vertical 
automorphisms, and how to identify them with the maps defined on a 
quantum structure group with values in $P$ and covariant under the 
adjoint coaction. All these results are in perfect correspondence with 
the classical situation, the only difference being that while in the 
case of a classical principal bundle gauge transformations may be 
viewed, via the above identifications,  as sections of the associated 
adjoint bundle, in the case of quantum bundles such an interpretation 
is impossible because a quantum adjoint bundle does not exist. 
 
In the classical case the above results are usually discussed in the 
context of locally trivial bundles and they are proved first for 
a trivial bundle and then deduced globally by patching 
trivial bundles together. Our experience in working with quantum fibre 
bundles tells us however that, although the notion of a locally 
trivial quantum bundle may be rigorously defined 
\cite{brzezinski6,pflaum1}, the resulting construction is not so 
natural as the classical one and usually leads to some technical 
difficulties. Therefore we prefer not to assume the local triviality 
of quantum bundles as long as possible and we employ a technique of 
proving of the above mentioned results that 
does not make use of the local structure of a bundle. The main tool 
that allows us to avoid the use of locally trivial bundles in this paper is 
a non-commutative generalisation of a translation map. 
 
Classically, a translation map is defined as follows. Assume that we 
have a manifold with a free action of a Lie group $G$. Every two 
points on an orbit  are then related by a unique element of $G$. A 
translation map assigns such an element of $G$ to any two points on an 
orbit. The notion of a principal bundle is equivalent to 
the existence of a continuous translation map 
\cite[Section~4.2]{husemoller1}. We construct the non-commutative 
version of a translation map by dualisation and show that the notion 
of a quantum principal bundle is equivalent to the existence of this 
generalised translation map. We then use this map throughout the paper 
to prove the above mentioned results. 
 
Our paper is organised as follows. In Section~2 we briefly summarise 
the basic facts about Hopf algebras and quantum 
bundles. In Section~3 we define 
a translation map in a quantum principal bundle and analyse some of 
its properties. Section~4 is devoted to analysis of cross sections of a 
quantum fibre bundle. In Section~5, we identify vertical 
automorphisms of a quantum principal bundle with the maps covariant 
under the adjoint coaction, and give various equivalent descriptions 
of gauge transformations of a trivial quantum principal bundle. The paper 
ends with a computation of the gauge group for examples of  bundles 
with  finite structure groups in Section~6. 
 
\section{Preliminaries} 
\subsection{Notation} 
\noindent Here we summarise the notation we use in the sequel. $A$ 
denotes a Hopf algebra over a field $k$ of real or complex numbers, 
with a coproduct $\Delta :A\to A\otimes A$, counit $\epsilon :A\to k$ 
and the antipode $\S: A\to A$ \cite{sweedler1}. For a 
coproduct we use an explicit  
expression $\Delta (a) = a\o\otimes a\t$, where the summation is 
implied according to the Sweedler sigma convention \cite{sweedler1}, i.e. 
$a\o\otimes a\t = \sum\sb{i\in I} a\o\sp i\otimes a\t\sp i$ for an 
index set $I$. We also  use the notation  
$$ 
a\o\tens a\t\tens\cdots\tens a\sb{(n)} = 
(\Delta\tens\underbrace{\id\tens\cdots\tens\id}\sb{n-2})\circ 
\cdots\circ(\Delta\tens\id)\circ\Delta  
$$ 
which describes a multiple action of $\Delta$ on $a\in A$. 
 
If $A$ is a Hopf algebra then $A\sp*$ denotes a dual Hopf algebra. 
$A\sp*$ has an algebra structure induced from the coalgebra structure 
of $A$ and a coalgebra structure induced from the algebra structure of 
$A$. For example $\forall x,y\in A\sp*$, $\forall a,b\in A$, $\langle 
xy,a\rangle = \langle x,a\o\rangle\langle y,a\t\rangle$, 
$\langle\Delta x,a\tens b\rangle = \langle x,ab\rangle$ etc.~, where 
$\langle\; ,\;\rangle :A\sp*\tens A\to k$ denotes the natural pairing. 
 
Recall that a vector space $V$ is called a {\em right $A$-comodule} if there 
exists a linear map $\rho\sb R: V\to V\tens A$, called a {\em right 
coaction}, such that $(\rho\sb  
R\tens\id)\circ\rho\sb R = (\id\tens\Delta)\circ\rho\sb R$ and 
$(\id\tens \epsilon )\circ\rho\sb R = \id$. Similarly, a vector 
space $V$ is called a {\em left $A$-comodule} if there 
exists a linear map $\rho\sb L: V\to A\tens V$, called a {\em left 
coaction}, such that $(\Delta\tens\id)\circ\rho\sb L = (\id\tens\rho\sb 
L)\circ\rho\sb L$ and 
$(\epsilon\tens \id )\circ\rho\sb L = \id$. 
We say that a unital algebra $P$ over $k$ is a {\em right A-comodule 
algebra} if $P$ is a right $A$-comodule with a coaction 
$\Delta\sb R :P\to P\otimes A$, and $\Delta\sb R$ is an algebra 
map.  The algebra structure of $P\tens A$ is that of a tensor product 
algebra. For a coaction $\Delta\sb R$ we use an explicit notation 
$\Delta\sb R u = u\sp{(\bar{1})}\otimes u\sp{(\bar{2})}$, where 
the summation is also implied. Notice that $u\sp{(\bar{1})}\in 
P$ and $u\sp{(\bar{2})}\in A$.

If $P$ is a right $A$-comodule algebra then $P\sp A$ denotes a fixed 
point subalgebra of $P$, i.e.  
$P\sp A = \{u\in P :\Delta\sb R u = u\otimes 1\}$.  
$P\sp A$ is a subalgebra of $P$ with a natural inclusion $j: 
P\sp A \hookrightarrow P$. In what follows we do  not write this 
inclusion explicitly but it should be understood that the elements of 
$P\sp A$ are viewed as elements of $P$ via $j$. 
 
Let $A$ be a Hopf algebra, $B$ be a unital algebra over $k$, and 
let $f,g :A\to B$ be linear maps. A {\em convolution product} of 
$f$ and $g$ is a linear map $f*g: A\to B$ given by 
$(f*g)(a) = f(a\o)g(a\t)$, for any $a\in A$. With respect to  
the convolution product, the set of all linear 
maps $A\to B$ forms an  
associative  algebra with the unit $1\sb B\epsilon$. We say that a 
linear map $f: A\to B$ is   
{\em convolution invertible} if there is a map $f\sp{-1}: A\to 
B$ such that 
$f*f\sp{-1} = f\sp{-1}*f = 1\sb B\epsilon$. The set of all convolution 
invertible maps $A\to B$ forms a multiplicative group. Similarly if 
$V$ is a right $A$-comodule and $f:V\to B$, $g:A\to B$ are linear maps 
then we define a convolution product $f*g:V\to B$ to be $(f*g)(v) = 
f(v\bo)g(v\bt)$. 
 
In this paper we work with a universal differential structure  
\cite{kastler1,kunz1}.  
 
\subsection{Quantum Fibre  Bundles} 
\noindent In this section we recall the basic elements of the theory 
of  
quantum principal and associated bundles \cite{brzezinski6}. 
 
Let $A$ be a Hopf algebra, $P$ a right $A$-comodule algebra 
with a coaction  
$\Delta\sb R :P\to P\otimes A$. We define a map $\chi :P\otimes P \to  
P\otimes A$, 
\begin{equation} 
\chi = (\cdot \otimes \id)\circ (\id \otimes \Delta\sb R). 
\label{chi} 
\end{equation} 
Explicitly, $\chi (u\otimes v) = uv\sp{(\bar{1})}\otimes 
v\sp{(\bar{2})}, $ for any $u,v\in P$. We say that the coaction 
$\Delta\sb R$ is  
{\elevenit free} if $\chi$ is a surjection. We define also a map 
$\;\;\tilde{}:P\sp 2 \to P\otimes\ker\epsilon$, by   
$ \tilde{} =  
\chi\mid\sb{P\sp 2}$. Here and below, for any algebra $P$,   
$P\sp 2\subset P\tens P$ denotes 
the kernel of the multiplication $\cdot$ in $P$. Let $B = P\sp A$.  
We say that the coaction $\Delta\sb R$ of $A$ on 
$P$ is {\elevenit exact}, if 
$\ker \;\tilde{}\; = P B\sp 2 P$.  
\begin{df}[\cite{brzezinski6}] 
\rm Let $A$ be a Hopf algebra, $(P ,\Delta\sb{R})$ be a right $A$-comodule 
algebra and let  $B = P\sp A$. We say that $P(B,A)$ is a {\em quantum 
principal bundle} with universal differential   
structure, with a structure quantum group $A$ and a base $B$ if the 
coaction $\Delta\sb R$ is free and exact. 
\label{df.principal} 
\end{df} 
 
The basic examples of  quantum principal bundles are the trivial bundle 
$P(B,A,\Phi)$ with trivialisation $\Phi:A\to P$ \cite[Example
4.2]{brzezinski6},   
and the bundle $P(B,A,\pi)$ \cite[Lemma 5.2]{brzezinski6}. In the latter  
case $P$ and $A$ are  
Hopf algebras and $\pi:  P\to A$ is a Hopf algebra projection, used 
in the construction of a quantum homogeneous space $B=P\sp A$. A large
number of  
examples of quantum principal bundles $P(B,A,\pi)$ has been found 
recently in \cite{meyer1} 
 
For a trivial quantum principal bundle $P(B,A,\Phi)$ one defines a  
{\em gauge transformation}  as a convolution invertible  map $\gamma: 
A\to B$ such that $\gamma(1) =1$. The set of all gauge transformations 
of $P(B,A,\Phi)$ forms a group with respect to the 
convolution product. This group is denoted by $\CA(B)$. A map
$\Psi:A\to P$ is a  
trivialisation of $P(B,A,\Phi)$ if and only if there exists
$\gamma\in\CA(B)$  
such that $\Psi =\gamma*\Phi$. 
 
\begin{df}[\cite{brzezinski6}] 
\rm Let $P(B,A)$ be a quantum principal bundle and let $V$ be a right 
$A^{\rm op}$-comodule algebra, where $A^{\rm op}$ denotes the algebra 
which is isomorphic to $A$ as a vector space but has an opposite 
product, with coaction $\rho_{R} : V \rightarrow V  
\otimes A$. The space $P \otimes V$ is naturally endowed with a right 
$A$-comodule structure  $\Delta_{E} : P \otimes V 
\rightarrow P \otimes V \otimes A$ given by 
$\Delta_{E} (u \otimes v) =  u^{(\overline 1)} \otimes 
v^{(\overline 1)} \otimes u^{(\overline 2)}v^{(\overline 2)}$ 
for any $u \in P$ and $v \in V$. We say that the fixed point subalgebra 
$E$ of $P\tens A$ with respect to $\Delta\sb E$  
is a  {\em quantum fibre bundle associated to $P(B,A)$} over $B$ with 
structure quantum group $A$ and standard fibre $V$. We denote it by $E = 
E(B,V,A)$.\footnote{A slightly different definition of $E(B,V,A)$ was
proposed  
in \cite{hajac1}. The formalism developed in this paper can be  
equally well applied to quantum fibre bundles in the sense of \cite{hajac1}.} 
\end{df}

\section{Definition and Properties of a Translation Map} 
In this section we give a definition and analyse transformation 
properties of a translation map in a quantum principal bundle. 
\begin{df} 
\rm Let $P(B,A)$ be a quantum principal bundle. A linear  map $\tau:A\to 
P\otimes{}\sb B P$,  given by $\tau(a) = \sum\sb{i\in I}u\sb  
i\otimes{}\sb B v\sb i$, where $\sum\sb{i\in I}u\sb i\otimes v\sb i \in 
\chi\sp{-1}(1\otimes a)$, is called a {\em translation map}. We will 
often use an explicit notation $\tau(a)= \tauo(a)\tens{}\sb 
B\taut(a)$.  
\label{definition.tau} 
\end{df} 
Since $\chi$ is a surjection, $\tau$ is defined on the whole of $A$. 
Moreover, if $a=0$ then, by exactness of the coaction, the 
corresponding $\sum\sb{i\in I}u\sb i\tens v\sb i \in PB\sp 2 P$. Hence 
$\sum\sb{i\in I}u\sb i\otimes{}\sb B v\sb i = 0$ and the map $\tau$ is 
well-defined as a linear map. In fact, a translation map of 
Definition~\ref{definition.tau} is well-defined if 
and only if $P$ is a total space of a quantum principal bundle.  
\begin{lemma} 
Let $P$ be a right $A$-comodule algebra with a free coaction 
$\Delta\sb R: P\to P\otimes A$. Let $B=P\sp A$. If there is a 
translation map $\tau :A\to P\tens{}\sb B P$ in $P$ then the coaction 
$\Delta\sb R$ is exact and hence there is a quantum principal bundle 
$P(B,A)$. 
\end{lemma} 
\proof We need to show that if $\sum\sb{i\in I}u\sb i\tens v\sb i \in 
\ker\;\tilde{}\;$ then $\sum\sb{i\in I}u\sb i\tens v\sb i \in PB\sp 2
P$. Take  
any $\sum\sb{i\in I}u\sb i\tens v\sb i \in 
\ker\;\tilde{}\;$ then $\sum\sb{i\in I}u\sb i\tens v\sb i \in 
\chi\sp{-1}(1\otimes 0)$. Since there is a translation map in $P$ we 
deduce that $\sum\sb{i\in I}u\sb i\otimes{}\sb B v\sb i = 0$, what implies 
that $\sum\sb{i\in I}u\sb i\tens v\sb i \in PB\sp 2 P$. \endproof 
 
Definition~\ref{definition.tau} of a translation map reproduces 
exactly the classical definition \cite[Definition~2.1]{husemoller1}, 
but in a language  
of algebras of functions on manifolds rather than manifolds themselves. 
Classically, if $X$ is a manifold on which a Lie group $G$ acts 
freely then the translation map $\hat{\tau}:X\times{}\sb MX\to G$, where 
$M=X/G$, is defined by $u\hat{\tau}(u,v)=v$. 
Dualising this construction we arrive immediately at 
Definition~\ref{definition.tau}.  
 
\begin{ex} 
In a trivial quantum principal bundle $P(B,A,\Phi)$ the translation 
map is given by 
\begin{equation} 
\tau(a) = \Phi\sp{-1}(a\o)\tens{}\sb B\Phi(a\t). 
\label{tau.trivial} 
\end{equation} 
\end{ex} 
\proof Using the fact that the trivialisation $\Phi$ is an intertwiner,  
i.e. $\Delta\sb R\Phi = (\Phi\otimes\id)\Delta$, and that $\Phi(1) 
= 1$ we obtain 
\begin{eqnarray*} 
\chi( \Phi\sp{-1}(a\o)\tens\Phi(a\t)) & = & 
\Phi\sp{-1}(a\o)\Phi(a\t)\bo\tens\Phi(a\t)\bt \\ 
& = & \Phi\sp{-1}(a\o)\Phi(a\t)\tens a\th = 1\tens a, 
\end{eqnarray*} 
for any $a\in A$. Hence the map $\tau$ given by 
Eq.~(\ref{tau.trivial}) is a translation map as stated. \endproof 
\begin{ex} 
In a quantum principal bundle $P(B,A,\pi)$ on a quantum homogeneous space 
$B$  the translation map $\tau: A\to 
P\otimes{}\sb B P$ is given by 
\begin{equation} 
\tau(a) = \S u\o\otimes{}\sb B u\t, 
\label{tau.homogeneous} 
\end{equation} 
where $u\in\pi\sp{-1}(a)$. 
\end{ex} 
\proof For any $a\in A$ we apply the map $\chi$ to $\S u\o\tens u\t$, 
where $u\in\pi\sp{-1}(a)$, to obtain 
$$ 
\chi(\S u\o\tens u\t) = (\S u\o) u\t\bo\tens u\t\bt  
 =  (\S u\o) u\t\tens \pi(u\th) = 1\tens\pi(u) = 1\tens a. 
$$ 
We conclude that $\tau$ given by Eq.~(\ref{tau.homogeneous}) is a 
translation map as stated. \endproof 
 
Before we analyse some properties of a translation map in a quantum 
principal bundle we study the behaviour of the map 
$\chi$, given by Eq.~(\ref{chi}), with respect to the coaction 
$\Delta\sb R$. Firstly we observe  
that if $P$ is a right $A$-comodule  then also $P\otimes P$ 
is a right $A$-comodule  with a coaction 
$ 
\deltens = (\id\otimes\id\otimes\cdot)\circ(\id\otimes\sigma\sb 
P\otimes\id)\circ(\Delta\sb R\otimes\Delta\sb R), 
$ 
where $\sigma\sb P:P\otimes P\to P\otimes P$ is a twist map $\sigma\sb 
P:u\otimes v \mapsto v\otimes u$. Explicitly, 
$ 
\deltens (u\otimes v) = u\bo\otimes v\bo\otimes u\bt v\bt . 
$ 
Secondly, both $P$ and $A$ are  right $A$-comodules with the coactions 
$\Delta\sb R: P\to P\otimes A$ and $\ad :A\to A\otimes 
A$,   
$ 
\ad(a) =  a\t\otimes (\S a\o)a\sb{(3)}. 
$ 
Therefore $P\otimes A$ is a right $A$-comodule with the coaction 
$$ 
\delad = (\id\otimes\id\otimes\cdot)\circ(\id\otimes\sigma\sb{PA} 
\otimes\id)\circ(\Delta\sb R\otimes\ad), 
$$ 
where  $\sigma\sb{PA}:P\otimes A\to A\otimes P$ is a twist map. 
 
Finally we define a linear map $\nu : P\tens A \to A\tens P\tens A$, 
$ 
\nu: u\tens a \mapsto u\bt\S a\o\tens u\bo\tens a\t . 
$ 
 Now we 
can prove the following lemma. 
\begin{lemma} 
Let $P$ be a right $A$-comodule algebra and let $\chi:P\otimes P \to  
P\otimes A$ be given by (\ref{chi}). Then 
  
1.  $(\id\tens\chi)\circ (\sigma\sb{PA}\circ\Delta\sb R\tens\id) = 
\nu\circ\chi$;  
 
2. $(\chi\tens\id)\circ(\id\tens\Delta\sb R) 
=(\id\tens\Delta)\circ\chi$;  
 
3. $(\chi\tens\id)\circ\deltens =\delad\circ\chi$. 
\label{lemma.chi} 
\end{lemma} 
\proof Since all the maps discussed in this lemma are linear it 
suffices to prove the required equalities for any $u\tens v \in P\tens 
P$. To prove the first assertion we compute 
$$ 
(\id\tens\chi)\circ (\sigma\sb{PA}\circ\Delta\sb R\tens\id)(u\tens v) 
 =  (\id\tens\chi)(u\bt\tens u\bo\tens v) = u\bt\tens u\bo v\bo\tens 
v\bt . 
$$ 
On the other hand 
\begin{eqnarray*} 
\nu\circ\chi(u\tens v) & = & \nu(uv\bo\tens v\bt)  
=  u\bt v\bt\o\S v\bt\t\tens u\bo v\bo \tens v\bt\th = u\bt\tens 
u\bo v\bo\tens v\bt . 
\end{eqnarray*} 
Thus 
$$ 
(\id\tens\chi)\circ (\sigma\sb{PA}\circ\Delta\sb R\tens\id) = 
\nu\circ\chi 
$$ 
and the first assertion of the lemma holds. 
 
The second assertion follows from the definition of $\chi$ and the 
fact that  $\Delta\sb R$ is a coaction.  Explicitly, 
\begin{eqnarray} 
(\chi\tens\id)\circ(\id\tens\Delta\sb R) & = & 
(\cdot\tens\id\tens\id)\circ(\id\tens\Delta\sb 
R\tens\id)\circ(\id\tens\Delta\sb R)\nonumber \\ 
& = & 
(\cdot\tens\id\tens\id)\circ(\id\tens\id\tens\Delta)\circ(\id\tens\Delta\sb 
R) \nonumber\\  
& = & (\id\tens\Delta)\circ(\cdot\tens\id)\circ(\id\tens\Delta\sb 
R) 
 =  (\id\tens\Delta)\circ\chi \nonumber 
\end{eqnarray} 
 
Finally, the third assertion was proven in  
\cite[Lemma~4.3]{brzezinski6}. It is also a consequence of the first 
two assertions. \endproof 
 
Now we can state the proposition that collects the transformation 
properties of a  
translation map in a quantum principal bundle. 
\begin{prop} 
The translation map $\tau: A\to 
P\otimes \sb B P$ in a quantum principal bundle $P(B,A)$ has the 
following properties: 
   
1. $(\sigma\sb{PA}\circ\Delta\sb R\otimes \sb B\id)\circ\tau = (\S 
\otimes\tau)\circ\Delta$;  
   
2. $(\id\otimes \sb B\Delta\sb R)\circ\tau = (\tau\otimes\id)\circ\Delta$; 
 
3. $\deltens\circ\tau = (\tau\otimes\id)\circ\ad $ 
 
4. $\cdot\circ\tau =1\epsilon$. 
\label{proposition.properties.phi} 
\end{prop} 
\proof  
1. Let $\tau(a) =\tauo(a)\tens\sb B \taut(a)$ for any $a\in A$. The 
first assertion of  
Lemma~\ref{lemma.chi} yields  
$$ 
(\id\tens\chi)\circ(\sigma\sb{PA}\circ\Delta\sb 
R\otimes\id)(\tauo(a)\tens\taut(a))   
= \S a\o\tens 1\tens a\t . 
$$ 
Since $\chi(\tauo(a)\tens\taut(a)) = 1\tens a$ for any $a\in A$ we 
immediately deduce that  
$$ 
(\sigma\sb{PA}\circ\Delta\sb R\otimes \sb B\id)\circ\tau = (\S 
\otimes\tau)\circ\Delta . 
$$ 
 
2. Using the second assertion of Lemma~\ref{lemma.chi} we obtain 
\begin{eqnarray*} 
(\id\tens{}\sb B\Delta\sb R)\circ \tau(a) & = & \tauo(a)\tens{}\sb B 
\taut(a)\bo \tens \taut(a)\bt \\ 
& = &  \tauo(a\o)\tens{}\sb B \taut(a\o) \tens a\t =   
\tau(a\o)\tens a\t , 
\end{eqnarray*} 
i.e.~the assertion. 
 
3. The third assertion of Lemma~\ref{lemma.chi} yields for any $a\in 
A$ 
\begin{eqnarray*} 
(\chi \tens\id) \circ \deltens(\tauo(a)\tens \taut(a)) & = & 
1\tens\ad(a) 
 =  1\tens a\t\tens\S a\o a\th. 
\end{eqnarray*} 
Hence, using the definition of $\tau$, we immediately find that 
$$ 
\Delta\sp\tens\sb R \circ\tau = (\tau\tens\id)\circ\ad . 
$$ 
 
4. For any $a\in A$ we have 
\begin{equation} 
  \tauo(a)\taut(a)\bo \tens \taut(a)\bt = 1\tens a. 
\label{chichi} 
\end{equation} 
Applying $\id\tens\eps$ to both sides of Eq.~(\ref{chichi}) we 
immediately obtain the assertion. This ends the proof of the 
proposition. \endproof 
 
\section{Cross Sections of a Quantum Fibre Bundle} 
In this section we use the notion of a translation map in a quantum 
principal bundle $P(B,A)$ to identify cross sections of a quantum 
fibre bundle $E(B,V,A)$  with equivariant maps $V\to P$. Recall that a 
linear map $\phi:V\to P$ is said to be equivariant if $\Delta\sb R 
\phi = (\phi\tens\id)\rho\sb R$, where $\rho\sb R$ is a right coaction 
of $A$ on $V$. In particular, our identification 
implies that a quantum principal bundle is trivial if it admits a 
cross section which is an algebra map. We begin with the following 
definition.  
\begin{df} 
\rm Let $E(B,V,A)$ be a quantum fibre bundle associated to a quantum 
principal bundle $P(B,A)$. A left $B$-module map 
$s: E\to B$ such that $s(1) = 1$ is called a {\em cross section} of 
$E(B,V,A)$. The set of cross sections of $E(B,V,A)$ is denoted by 
$\Gamma(E)$.  
\label{definition.section} 
\end{df} 
\begin{lemma} 
If $s:E\to B$ is a cross section of a quantum fibre bundle $E(B,V,A)$  then 
$s\circ j\sb E = \id$, where $j\sb E :B\hookrightarrow E$ is a natural  
inclusion $j\sb E:b\mapsto b\tens 1\sb V$. 
\label{lemma.section} 
\end{lemma} 
\proof For any $b\in B$, $s\circ j\sb E(b) = s(b\tens 1) = bs(1) =b$. 
\endproof  
 
The result of trivial Lemma~\ref{lemma.section} justifies the term 
cross section used in Definition~\ref{definition.section}. We remark that in 
\cite{brzezinski6} cross sections of a quantum fibre bundle were 
defined as maps  
$E\to B$ having the property described in Lemma~\ref{lemma.section}. 
Definition~\ref{definition.section} is more restrictive than that of 
 \cite{brzezinski6} since the fact that $s\circ j\sb E=\id$ does not 
imply that $s$ is a left $B$-module map. We also remark that the  
definition of a cross section of a quantum fibre bundle analogous to the 
one we use 
here was first proposed  in \cite{hajac1}.

Now we can state the first of two main results of this section. 
\begin{thm} 
Let $A$ be a Hopf algebra with a bijective antipode. Cross sections of 
a quantum fibre bundle $E(B,V,A)$ associated to a  
quantum principal bundle $P(B,A)$ are in bijective 
correspondence  
with  equivariant maps $\phi :V\to P$ such that $\phi(1) = 1$.\nolinebreak 
\label{proposition.section.vector} 
\end{thm} 
\proof The fact that each equivariant map $\phi :V \to P$ induces a 
map $s$ such that $s\circ j\sb E =\id$,  by $s = 
\cdot(\id\tens\phi)$ was proved in 
\cite[Proposition~A.5]{brzezinski6}. Clearly, $s(bx) = bs(x)$ for any 
$b\in B$ and $x\in E$, and s(1) = 1, hence $s$ defined above is a 
cross section of $E(B,V,A)$. 
 
Conversely, for any $s\in\Gamma(E)$ we define a map $\phi: V\to P$ by 
\begin{equation} 
\phi : v\mapsto \tauo(\S\sp{-1}v\bt)s(\taut(\S\sp{-1}v\bt)\tens v\bo), 
\label{phi.s.def} 
\end{equation} 
where $\tau(a) = \tauo(a)\tens{}\sb B\taut(a)$ is a translation map in 
$P(B,A)$. We observe that this definition of $\phi$ makes sense since 
$s$ is a $B$-module map and, by 
Proposition~\ref{proposition.properties.phi}, $\forall v\in V$, 
$$ 
\tau(\S\sp{-1}v\bt)\tens v\bo\in P\tens{}\sb BE. 
$$ 
Explicitly, 
\begin{eqnarray*} 
(\id\tens{}\sb B\Delta\sb E) (\tau(\S\sp{-1}v\bt)\tens v\bo) & = &  
\tauo(\S\sp{-1}v\bt\t)\tens{}\sb B \taut(\S\sp{-1}v\bt\t)\bo\tens 
v\bo \tens\\ 
&&\tens \taut(\S\sp{-1}v\bt\t)\bt v\bt\o \\ 
& = & \tauo(\S\sp{-1}v\bt\th)\tens{}\sb B \taut(\S\sp{-1}v\bt\th)\tens 
v\bo\tens\\ 
&&\tens (\S\sp{-1}v\bt\t) v\bt\o \\ 
& = & \tau(\S\sp{-1}v\bt)\tens v\bo\tens 1. 
\end{eqnarray*} 
Clearly, $\phi(1) = 1$. Furthermore, using 
Proposition~\ref{proposition.properties.phi} we find that for any 
$v\in V$ 
\begin{eqnarray*} 
\Delta\sb R \phi(v) & = & \Delta\sb R 
(\tauo(\S\sp{-1}v\bt)s(\taut(\S\sp{-1}v\bt)\tens v\bo) \\ 
& = & (\tauo(\S\sp{-1}v\bt)\bo s(\taut(\S\sp{-1}v\bt)\tens v\bo)\tens 
\tauo(\S\sp{-1}v\bt)\bt \\ 
& = & (\tauo(\S\sp{-1}v\bt\o)\bo s(\taut(\S\sp{-1}v\bt\o)\tens v\bo)\tens 
\S\circ\S\sp{-1}v\bt\t \\ 
& = & \phi(v\bo)\tens v\bt = (\phi\tens\id)\circ\rho\sb R , 
\end{eqnarray*} 
hence $\phi$ is an equivariant map as required. 
 
Therefore we have constructed the maps $\theta
:\phi\mapsto\cdot(\id\tens\phi)$  
and $\tilde{\theta}:s\mapsto \phi$, where $\phi$ is given by 
Eq.~(\ref{phi.s.def}). We now show  
that they are inverses to each other. For any $s\in\Gamma(E)$ and 
$\sum\sb{i\in I}u\sb i\tens v\sb i\in E$ we have 
$$ 
(\theta\circ\tilde{\theta})(s)(\sum\sb{i\in I}u\sb i\tens v\sb i) 
=\sum\sb{i\in I} u\sb i\tilde{\theta}(s)(v\sb i) = 
\sum\sb{i\in I}u\sb i\tauo(\S\sp{-1} v\sb i\bt)s(\taut(\S\sp{-1} v\sb 
i\bt)\tens v\sb i\bo). 
$$ 
Further, using Proposition~\ref{proposition.properties.phi} we 
find 
\begin{eqnarray*} 
&&\hspace{-.4cm}(\sigma\sb{PA}\circ\Delta\sb R \tens{}\sb 
B\id\tens\id)(\sum\sb{i\in I}u\sb i\tauo(\S\sp{-1} v\sb 
i\bt)\tens{}\sb B\taut(\S\sp{-1}  
v\sb i\bt)\tens v\sb i\bo)\\ 
&&\hspace{.15cm} = \sum\sb{i\in I}u\sb i\bt v\sb i\bt\t\tens u\sb i\bo 
\tauo(\S\sp{-1}  
v\sb i\bt\o)\tens{}\sb B\taut(\S\sp{-1}  
v\sb i\bt\o)\tens v\sb i\bo \\ 
&&\hspace{.15cm} = \id\tens(\cdot\tens{}\sb 
B\tens\id\tens\id)\circ(\id\tens\tau\circ\S\sp{-1}\tens\id) 
\circ(\id\tens\sigma\sb{VA}\circ\rho\sb  
R)(\sum\sb{i\in I}u\sb i\bt v\sb i\bt\tens u\sb i\bo\tens v\sb i\bo). 
\end{eqnarray*} 
Since $\sum\sb{i\in I}u\sb i\tens v\sb i\in E$ we obtain 
\begin{eqnarray*} 
&&(\sigma\sb{PA}\circ\Delta\sb R \tens{}\sb 
B\id\tens\id)(\sum\sb{i\in I}u\sb i\tauo(\S\sp{-1} v\sb 
i\bt)\tens{}\sb B\taut(\S\sp{-1}  
v\sb i\bt)\tens v\sb i\bo )\\  
&&\hspace{2cm} =  \id\tens(\cdot\tens{}\sb 
B\tens\id\tens\id)\circ(\id\tens\tau\circ\S\sp{-1}\tens\id) 
\circ(\id\tens\sigma\sb{VA}\circ\rho\sb  
R)(\sum\sb{i\in I}1\tens u\sb i\tens v\sb i) \\ 
&&\hspace{2cm} = \sum\sb{i\in I}1\tens u\sb i\tauo(\S\sp{-1} v\sb 
i\bt)\tens{}\sb B\taut(\S\sp{-1}  
v\sb i\bt)\tens v\sb i\bo. 
\end{eqnarray*} 
Hence 
$$ 
\sum\sb{i\in I}u\sb i\tauo(\S\sp{-1} v\sb i\bt)\tens{}\sb B\taut(\S\sp{-1} 
v\sb i\bt)\tens v\sb i\bo \in B\tens{}\sb B E. 
$$ 
Therefore using the fact that $s$ is a left $B$-module map and 
Proposition~\ref{proposition.properties.phi} again, we obtain 
$$ 
(\theta\circ\tilde{\theta})(s)(\sum\sb{i\in I}u\sb i\tens v\sb i) = 
s(u\sb i\tauo(\S\sp{-1} 
v\sb i\bt)\taut(\S\sp{-1} v\sb i\bt)\tens v\sb i\bo) = s(\sum\sb{i\in 
I}u\sb i\tens v\sb i). 
$$ 
Conversely, for any $v\in V$ and equivariant $\phi :V\to P$ 
\begin{eqnarray*} 
(\tilde{\theta}\circ\theta)(\phi)(v) & = & 
\tauo(\S\sp{-1}v\bt)\theta(\phi)(\taut(\S\sp{-1}v\bt)\tens v\bo) \\ 
& = &\tauo(\S\sp{-1}v\bt)\taut(\S\sp{-1}v\bt)\phi(v\bo) = \phi(v). 
\end{eqnarray*} 
\endproof 
 
\begin{rk} 
\rm There is a certain class of quantum fibre bundles  
\cite[Definition 11]{budzynski1} for which  
Theorem~\ref{proposition.section.vector} holds  even if  
the antipode $\S:A\to A$ is not bijective.   
We consider a 
{\em left} $A$-comodule algebra with a coaction $\rho\sb L: V\to 
A\tens V$. We view $V$ as a right $A\sp{op}$-comodule algebra with a right 
coaction $\rho\sb R = (\id\tens\S)\circ\sigma\sb{AV}\circ\rho\sb L$, 
where $\sigma\sb{AV} : A\tens V\to V\tens A$ is a twist map, and 
consider a quantum fibre bundle $E(B,V,A)$ associated to $P(B,A)$.  
In the case of such a bundle, to each 
cross-section $s$ we associate a map $\phi :V\to P$ given by 
$\phi=(\cdot\tens\id)\circ(\id\tens \sb 
Bs)\circ(\tau\tens\id)\circ\rho\sb L$ and proceed  
as in the proof of Theorem~\ref{proposition.section.vector} to show 
that $s\mapsto \phi$ establishes the required bijective correspondence. 
\label{remark.def.trivial.bundle}\dia 
\end{rk} 
\begin{ex} 
\rm Let $E(B,V,A)$ be a quantum fibre bundle associated to a trivial 
quantum principal bundle $P(B,A,\Phi)$. In this case every element of 
$E$ has the from $\sum\sb{i\in I}b\sb i\Phi\sb E(v\sb i)$, where 
$b\sb i\in B$ and $v\sb i \in V$, and $\Phi\sb E: V\to E$, $\Phi\sb E: 
v\mapsto \Phi(\S\sp{-1}v\bt)\tens v\bo$ \cite[Appendix]{brzezinski6}. 
The isomorphisms $\theta$ and $\theta\sp{-1}$ of the 
proof of Theorem~\ref{proposition.section.vector} read 
$$ 
\theta(\phi)(\sum\sb{i\in I}b\sb i\Phi\sb E(v\sb i)) = \sum\sb{i\in 
I}b\sb i\Phi(\S\sp{-1}v\sb i\bt)\phi(v\sb i\bo), \quad 
\theta\sp{-1}(s)(v) = \Phi\sp{-1}(\S\sp{-1}v\bt)s(\Phi\sb E(v\bo)) 
$$ 
for any equivariant $\phi:V\to P$ and $s\in\Gamma(E)$. 
Notice that the map $\theta\sp{-1}(s)$ 
obtained in this way is different from the equivariant map $\phi$ 
discussed in \cite[Proposition A6]{brzezinski6}.

If $E(B,V,A)$ is of the type described in  
Remark~\ref{remark.def.trivial.bundle} then $\S\sp{-1}$ 
disappears from definitions of $\theta$ and $\theta\sp{-1}$. 
Explicitly we have 
$$ 
\theta(\phi)(\sum\sb{i\in I}b\sb i\Phi\sb E(v\sb i)) = \sum\sb{i\in 
I}b\sb i (\Phi*\phi)(v\sb i), \qquad\theta\sp{-1}(s) = 
\Phi\sp{-1}*(s\circ\Phi\sb E),  
$$ 
where $*$ denotes convolution product between the maps $f:A\to P$, $g: 
V\to P$, given by $f*g = \cdot\circ(f\tens g)\circ\rho\sb L$.\dia 
\end{ex} 
 
Before we describe an important corollary of  
Theorem~\ref{proposition.section.vector}, we state a lemma that allows 
one to view a quantum principal bundle as a quantum fibre bundle. 
\begin{lemma} 
A quantum principal bundle $P(B,A)$ is a fibre bundle associated to 
$P(B,A)$ with the fibre which is isomorphic to $A$ as an algebra and 
with the coaction $\rho\sb R=(\id\otimes \S)\circ\Delta'$, where 
$\Delta'$ denotes the opposite coproduct, $\Delta'(a) = a\t\tens a\o$. 
\label{example.vector.principal} 
\end{lemma} 
\proof Firstly we show that the total space $E$ of the bundle 
constructed in the lemma is equal to $\im\Delta\sb R$. We take any 
$u\bo\tens u\bt\in \im\Delta\sb R$. Then 
$$ 
\Delta\sb E(u\bo\tens u\bt) = u\bo\tens u\bt\th\tens u\bt\o\S u\bt\t 
= u\bo\tens u\bt\tens 1, 
$$ 
hence $u\bo\tens u\bt\in E$. Conversely, let $\sum\sb{i\in I} 
u\sb i\tens a\sp i \in E$. Then 
$$ 
\sum\sb{i\in I} u\sb i\bo\tens a\sp i\t\tens u\sb i\bt\S a\sp 
i\o = \sum\sb{i\in I} u\sb i\tens a\sp i\tens 1. 
$$ 
Applying $\id\tens\cdot\circ\sigma\sb A$, where $\sigma\sb A: A\tens 
A\to A\tens A$ is a twist map, to both sides of the above 
equation we obtain 
$$ 
\sum\sb{i\in I} u\sb i\tens a\sp i = \sum\sb{i\in I}\eps(a\sp 
i)u\sb i\bo\tens u\sb i\bt, 
$$ 
hence $\sum\sb{i\in I} u\sb i\tens a\sp i$ manifestly lies in 
$\im\Delta\sb R$. 
 
Secondly, we observe that $\im\Delta\sb R \cong P$ as  
algebras, the isomorphism being provided by the coaction $\Delta\sb
R$. Clearly  
$\Delta\sb R :P\to \im\Delta\sb R$ is a surjection. Moreover, if 
$\Delta\sb Ru = 0$ then $u = (\id\tens\eps)\Delta\sb R u =0$, hence 
$\ker\Delta\sb R = \{ 0\}$. Since $\Delta\sb R$ is an algebra map, the 
isomorphism is established. Therefore we may identify $E$ with $P$ and 
$P(B,A)$ is a quantum fibre bundle as stated. \endproof 
 
\begin{cor} 
Cross sections $s:P\to B$ of a quantum principal bundle $P(B,A)$ are in 
bijective correspondence  
with the maps $\phi :A\to P$ such that $\Delta\sb R\phi = 
(\phi\tens\S)\Delta'$ and $\phi(1) = 1$. 
\label{proposition.section} 
\end{cor} 
\proof We identify $P(B,A)$ with a quantum fibre bundle $E(B,A,A)$ of 
Lemma~\ref{example.vector.principal} and then apply 
Theorem~\ref{proposition.section.vector}. When the total space $E 
=\im\Delta\sb R$ is viewed back as $P$ then the bijective maps 
$\theta$, $\tilde{\theta}$ come out as $\theta :\phi\mapsto\id*\phi$ 
and $\tilde{\theta}:s\mapsto \cdot(\id\tens{}\sb Bs)\tau$. By this 
identification we see that we need not assume  that the antipode is 
bijective since S does not enter the definition of $\tilde{\theta}$. 
On the other hand it also follows from the fact that $P(B,A)$ is a 
fibre bundle of the type described in 
Remark~\ref{remark.def.trivial.bundle}. One can  
also check directly that $\theta$, $\tilde{\theta}$ provide the 
required correspondence. \endproof 
 
We notice  that if $A$  has a  bijective antipode $\S$, the 
sections of  
a quantum principal bundle $P(B,A)$ are in one-to-one correspondence 
with the maps $\psi:A\to P$ such that $\psi(1) = 1$ and $\Delta\sb 
R\circ\psi = (\id\tens\psi)\circ\Delta$. We simply need to define 
$\psi=\phi\circ\S\sp{-1}$, where $\phi$ is given by 
Corollary~\ref{proposition.section}.  
\begin{prop} 
Any trivial quantum principal bundle $P(B,A,\Phi)$ admits a section. 
Conversely, if a bundle $P(B,A)$ admits a section which is an algebra 
map then $P(B,A)$ is trivial with the total space $P$ isomorphic to 
$B\tens A$ as an algebra. 
\label{prop.section.trivial} 
\end{prop} 
\proof A convolution inverse of a trivialisation $\Phi$ of a trivial quantum 
principal bundle  
$P(B,A,\Phi)$ satisfies the assumptions of 
Corollary~\ref{proposition.section}, hence $s=\id*\Phi\sp{-1}$ is a 
section of $P(B,A,\Phi)$. Conversely, assume that an algebra map 
$s:P\to B$ is a section of $P(B,A)$. Clearly, $s$ is a $B$-bimodule 
map, hence we can define a linear map $\Phi:A\to P$, $\Phi 
=\cdot(s\otimes{}\sb B\id)\circ\tau$. Let $\phi$ be a map 
$\tilde{\theta}(s)$ constructed in 
Corollary~\ref{proposition.section}. We will show that $\Phi$ and 
$\phi$ are convolution inverses to each other. It will then follow that 
$\Phi$ is a trivialisation of $P(B,A)$. We have 
\begin{eqnarray*} 
\Phi(a\o)\phi(a\t) & = & 
s(\tauo(a\o))\underbrace{\taut(a\o)\tauos(a\t)}s(\tauts(a\t)) \\ 
& = & s(\tauo(a\o))s(\taut(a\o)\tauos(a\t)\tauts(a\t))\\ 
& = & s(\tauo(a))s(\taut(a)) = s (\tauo(a)\taut(a)) = \eps(a). 
\end{eqnarray*} 
The symbol $\tauos(a)\otimes \tauts(a)$ here denotes the second copy of 
$\tau(a)$ and we have used  
Proposition~\ref{proposition.properties.phi} to  
deduce that the expression in the brace is in $B$. Similarly, 
\begin{eqnarray*} 
\phi(a\o)\Phi(a\t) & = & 
\tauo(a\o)s(\taut(a\o))s(\tauos(a\t))\tauts(a\t) \\ 
& = & \tauo(a\o)s(\underbrace{\taut(a\o)\tauos(a\t)})\tauts(a\t)\\ 
& = & \tauo(a\o)\taut(a\o)\tauos(a\t)\tauts(a\t)= \eps(a). 
\end{eqnarray*} 
As before we have used 
Proposition~\ref{proposition.properties.phi} to   
deduce that the  expression in the brace is in $B$. 
 
To prove that $P\cong B\tens A$ as algebras we consider a map $\Theta 
:P\to B\tens A$, $\Theta = (s\tens\id)\circ\Delta\sb R$. Clearly, 
$\Theta$ is an algebra map since both $s$ and $\Delta\sb R$ are 
algebra maps. Moreover the map $\tilde{\Theta} : P \to B\tens A$, 
$\tilde{\Theta} : b\tens a\mapsto bs(\tauo(a))\taut(a)$ is an inverse 
of $\Theta$. Explicitly, 
$$ 
\Theta\circ\tilde{\Theta}(b\tens a) = s(bs(\tauo(a\o))\taut(a\o))\tens 
a\t = bs(\tauo(a\o))s(\taut(a\o))\tens a\t = b\tens a, 
$$ 
and 
\begin{eqnarray*} 
\tilde{\Theta}\circ\Theta (u) & = & s(u\bo)s(\tauo(u\bt))\taut(u\bt) = 
s(u\bo\tauo(u\bt))\taut(u\bt)\\ 
&& = u\bo\tauo(u\bt)\taut(u\bt) = u.\qquad \Box 
\end{eqnarray*} 
 
Therefore we have obtained the criterion of triviality of a quantum 
principal bundle $P(B,A)$ which naturally generalises the classical 
case. In the classical limit all algebras are assumed commutative and 
all maps are algebra maps, hence 
Proposition~\ref{prop.section.trivial} states simply that a classical 
principal bundle is trivial if and only if it admits a  cross section.  
 
We also remark that the criterion of triviality of a quantum principal 
bundle similar to the one in Proposition~\ref{prop.section.trivial} 
was proved in \cite[Theorem~2]{budzynski1} in the case  of locally trivial 
bundles. The locally trivial bundles used there have total spaces 
locally isomorphic to the tensor product algebras $B\tens A$. We see 
that using the notion of a translation map we need not assume that a 
quantum principal bundle is locally trivial to prove 
Proposition~\ref{prop.section.trivial}. This fact reflects precisely 
the classical case in which the local triviality of a principal bundle 
is not necessary for validity of a classical version of 
Proposition~\ref{prop.section.trivial} \cite[Section~4.8]{husemoller1}. 
\begin{rk} 
\rm We would like to emphasise that the existence of a cross section of a 
quantum principal bundle does not necessarily imply that the bundle 
is trivial. As an example of a non-trivial quantum principal bundle 
admitting a cross section we consider the quantum Hopf fibration $SU\sb 
q(2)(S\sb q\sp 2,k[Z,Z\sp{-1}],\pi)$ \cite[Section~5.2]{brzezinski6}. The 
total space of this bundle is the quantum group $SU\sb 
q(2)$, as an algebra generated by the identity and a matrix 
$T = (t\sb{ij}) = \pmatrix{\alpha & \beta \cr \gamma &\delta}$ \cite{frt}. 
The base space $S\sb q\sp 2\subset SU\sb q(2)$ is a quantum two-sphere 
\cite{podles1}, defined as a fixed point subalgebra, $S\sb q\sp 2 = SU\sb 
q(2)\sp{k[Z,Z\sp{-1}]}$.  
 
It was shown in \cite{brzezinski6} that $SU\sb 
q(2)(S\sb q\sp 2,k[Z,Z\sp{-1}],\pi)$ is a non-trivial quantum 
principal bundle on the quantum homogeneous space $S\sb q\sp 2$.  
We consider a linear map 
$\phi : k[Z,Z\sp{-1}]\to SU\sb q(2)$, given by  
$$ 
\phi(1) = 1,\qquad \phi(Z\sp n) = \delta\sp n, \qquad \phi(Z\sp{-n}) 
=\alpha\sp n, 
$$ 
for any positive integer $n$. The map $\phi$ satisfies the hypothesis 
of Corollary~\ref{proposition.section}, hence it induces a cross section $s: 
SU\sb q(2) \to S\sb q\sp 2$, $s: u\mapsto u\o\phi(\pi(u\t))$.  
One can clearly check, however, that $s$ is not an algebra map. 
 
Finally we notice that in the semi-classical limit, $q\to 1$, the quantum 
principal  
bundle $SU\sb q(2)(S\sb q\sp 2,k[Z,Z\sp{-1}],\pi)$ reduces to the 
classical Hopf fibration written in terms of algebras of functions on 
manifolds. Also in this case the map 
$\phi$ above can  
be defined and hence the classical Hopf bundle admits a cross  section 
$s$ in the 
sense of Definition~\ref{definition.section}. But since $s$ is not an 
algebra map, the cross section obtained in this way is not continuous.\dia 
\end{rk} 
 
\section{Vertical Automorphisms of a Quantum Principal Bundle} 
\begin{df} 
\rm Let $P(B,A)$ be a quantum principal bundle. Any left $B$-module 
automorphism $\CF : P\to P$ such that $\CF (1) =1$ and $\Delta\sb R 
\CF = (\CF\otimes \id)\Delta\sb R$ is called a {\em vertical 
automorphism} of the 
bundle $P(B,A)$. The set of all vertical automorphisms of $P(B,A)$ is 
denoted by $Aut\sb B(P)$.  
\end{df} 
 
Elements of $Aut\sb B(P)$ preserve both the base space $B$ and the 
action of the structure quantum group $A$ of a quantum principal 
bundle $P(B,A)$. $ Aut\sb B(P)$ can be equipped with a multiplicative 
group structure $\cdot :(\CF\sb 1, \CF\sb 2)\mapsto \CF\sb 
2\circ\CF\sb 1$. Vertical automorphisms are often called gauge 
transformations and $ Aut\sb B(P)$ is termed a gauge group.   
\begin{prop} 
\begin{sloppy} Vertical automorphisms 
of a quantum principal bundle $P(B,A)$ are in bijective correspondence 
with convolution invertible  
maps $f:A\to P$ such that $f(1)=1$ and $\Delta\sb Rf = (f\otimes 
\id)\ad$.\end{sloppy} 
\label{prop.auto} 
\end{prop}

\proof Let $f$ be a map satisfying the hypothesis of the proposition. 
Define a map $\CF :P\to P$ by $\CF =\id*f$. We show now that the map 
$\CF$  is a vertical automorphism. First we need to prove 
that $\CF$ is a left $B$-module map. Take any $b\in B$ and $u\in P$. Then    
$$\CF(bu) =  bu\bo f(u\bt) =b\CF(u),$$ 
hence the map $\CF$ is a left $B$-module map as stated. To show that 
$\CF$ is right-invariant we take any $u\in P$ and compute  
\begin{eqnarray} 
\Delta\sb R\CF (u) & = & \Delta\sb R(  u\bo f(u\bt)) \nonumber  
=  u\bo f(u\bt\th)\otimes u\bt\o(\S u\bt\t)u\bt\f \nonumber \\ 
& = &   u\bo f(u\bt\o)\otimes u\bt\t = (\CF\otimes\id)\Delta\sb R 
u. \nonumber  
\end{eqnarray} 
Finally we have to show that $\CF$ is invertible. Consider a map 
$\tilde{\CF} = \id*f\sp{-1}$. First we observe that    
$ 
\Delta\sb R f\sp{-1} = (f\sp{-1}\otimes\id)\ad . 
$ 
This implies that $\tilde{\CF}$ is right-invariant. It is also 
a left $B$-module homomorphism. Now we use the right invariance of 
$\tilde{\CF}$ to compute   
$$ 
\CF\circ\tilde{\CF} = \tilde{\CF}* f = \id*f\sp{-1}*f = \id . 
$$ 
Similarly, by the right invariance of $\CF$ we obtain  
$ 
\tilde{\CF}\circ\CF = \id . 
$ 
Therefore the map $\tilde{\CF}$ is an inverse of $\CF$ and the first 
part of the proposition is proven. 
 
Conversely, for any $\CF\in Aut\sb B(P)$ we define a map 
$f: A\to P$,  
\begin{equation} 
f =\cdot\circ(\id\tens{}\sb B\CF)\circ\tau ,  
\label{map.f} 
\end{equation} 
where $\tau$ is a translation map. Explicitly  
$f(a) = \tauo(a)\CF(\taut(a))$.  
The map $f$ is well-defined since $\CF$ is a left $B$-module map.  
We need to show that $f$ given by Eq.~(\ref{map.f}) satisfies the 
assertion of the proposition. Clearly, $f(1) =1$. Next we derive the 
covariance property of $f$. We compute 
\begin{eqnarray*} 
\Delta\sb R f & = & \Delta\sb R\circ \cdot\circ(\id\tens{}\sb 
B\CF)\circ\tau  
= (\cdot\tens\id)\circ \deltens\circ (\id\tens{}\sb 
B\CF)\circ\tau \\ 
& = & (\cdot\tens\id)\circ (\id\tens{}\sb 
B\CF\tens\id)\circ\deltens\circ \tau  
= (\cdot\circ(\id\tens{}\sb B\CF)\circ\tau \tens\id)\circ\ad = 
(f\tens\id)\ad , 
\end{eqnarray*} 
where in the third equality we used that $\CF$ is an intertwiner, and 
in the fourth one we used the assertion 3 of 
Proposition~\ref{proposition.properties.phi}. 
 
 Consider a map 
$\tilde{f}:A\to P$, given by Eq.~(\ref{map.f}) but with $\CF$ replaced 
by its inverse $\CF\sp{-1}$. We show that the map $\tilde{f}$ is a 
convolution inverse of $f$. We have 
$$ 
\tilde{f}(a\o)f(a\t) = 
\tauo(a\o)\underbrace{\CF\sp{-1}(\taut(a\o))\tauos(a\t)}\CF(\tauts(a\t)). 
$$ 
Using the assertions 1 and 2  of 
Proposition~\ref{proposition.properties.phi} we can easily  
see that the expression in the brace is in $B$. Since $\CF$ is a 
left $B$-module map we obtain 
\begin{eqnarray*} 
\tilde{f}(a\o)f(a\t) &=& 
\tauo(a\o)\CF\left(\CF\sp{-1}(\taut(a\o))\tauos(a\t)\tauts(a\t)\right) \\ 
& = &  \tauo(a)\taut(a) = \eps(a). 
\end{eqnarray*} 
To derive the second and third equalities we used the assertion 4 of 
Proposition~\ref{proposition.properties.phi}. Similarly one proves 
that $f*\tilde{f}=\eps$.  
 
Therefore we have established the correspondence between the vertical 
automorphisms and normalised, convolution invertible, $\ad$-covariant 
maps $f:A\to P$. We need to prove that this correspondence is 
bijective. Denote by $\theta\sb A$ the map $f\mapsto \id*f$, and by 
$\tilde{\theta}\sb A$ the map $\CF\mapsto f$, where $f$ is given 
by Eq.~(\ref{map.f}). We will show that $\tilde{\theta}\sb A$ is an 
inverse of $\theta\sb A$. We have 
$$ 
\tilde{\theta}\sb A\circ \theta\sb A (f(a)) = \tilde{\theta}\sb 
A(\id*f)(a) = \tauo(a) (\id* f)(\taut(a)), 
$$ 
where $\tauo(a)\tens \taut(a)\in\chi\sp{-1}(1\tens a)$. But since $ 
\tauo(a)\taut(a)\bo\tens \taut(a)\bt  = 1\tens a$, we obtain 
$$ 
\tauo(a) (\id* f)(\taut(a)) =  
\tauo(a)\taut (a)\bo f(\taut(a)\bt) = f(a), 
$$ 
hence $\tilde{\theta}\sb A\circ \theta\sb A (f) =f$. 
 
Conversely, take any $\CF\in Aut\sb B(P)$. Then 
$$ 
\theta\sb A \circ \tilde{\theta}\sb A (\CF(u)) =  \underbrace{u\bo 
\tauo(u\bt)}\CF(\taut(u\bt)),  
$$ 
for any $u\in P$.  Using Proposition~\ref{proposition.properties.phi} 
we see that the expression in the brace 
is in $B$, hence, because $\CF$ is a left $B$-module map, 
$$ 
\theta\sb A \circ \tilde{\theta}\sb A (\CF(u)) = \CF(u\bo \tauo(u\bt) 
\taut(u\bt)) =  \CF(u\bo\eps(u\bt)) =\CF(u). 
$$ 
Therefore the map $\theta\sb A$ has an inverse and 
the proposition is proven. \endproof  
 
It is easily seen from the proof of  
Proposition~\ref{prop.auto} that  maps $f:A\to P$ form a 
group with respect to the convolution product. This group is 
denoted by $\CA(P)$.  In the classical case 
Proposition~\ref{prop.auto} allows one to interpret vertical 
automorphisms as cross sections of an associated adjoint bundle. This is 
because the elements of $\CA(P)$ are in bijective correspondence with 
such cross sections by  
Proposition~\ref{proposition.section}. In the general, 
non-commutative, situation there 
is no associated adjoint bundle and hence such an interpretation of 
vertical automorphisms is not possible. Still, as in the case of 
ordinary principal bundles,  
Proposition~\ref{prop.auto} implies the following: 
\begin{cor} 
$Aut\sb B(P)\cong\CA(P)$ as 
multiplicative groups. 
\label{cor.auto.gauge} 
\end{cor} 
\proof In the proof of Proposition~\ref{prop.auto} we have  
defined the bijective map $\theta\sb A: \CA(P)\to Aut\sb B(P)$, $\theta\sb A: 
f\mapsto \id*f$. We need to show that $\theta\sb A$ is a 
group homomorphism.   
We take any $f\sb 1, f\sb 2\in \CA(P)$ and compute     
$$ 
\theta\sb A(f\sb 1*f\sb 2)  =  \id *f\sb 1 *f\sb 2 =  \id (\theta\sb 
A(f\sb 1))*f\sb 2 = \theta\sb A(f\sb 2)\circ \theta\sb A(f\sb 1) = 
\theta\sb A(f\sb 1)\theta\sb A(f\sb 2). 
$$ 
From the proof of Proposition~\ref{prop.auto} it is clear that 
$\theta\sb A(f\sp{-1}) = \theta\sb A(f)\sp{-1}$. Furthermore, 
$\theta\sb A(1\eps) = \id*\eps = \id$, hence the map 
$\theta\sb A$ is a group homomorphism. \endproof

Now we examine some  properties of vertical automorphisms in the case 
of a trivial quantum principal bundle. In particular we identify 
vertical automorphisms of $P(B,A,\Phi)$ with gauge transformations. 
\begin{thm}  
Let $P(B,A,\Phi)$ be a trivial quantum principal bundle. Then the 
groups $Aut\sb B(P)$, $\CA(P)$, and the gauge group 
$\CA(B)$ are isomorphic to each other.  
If $A$ and $B$ are finite dimensional then the above groups are 
isomorphic to the group $\CR(B)$ of all invertible elements $v\in 
B\otimes A\sp*$ such that $(\id\otimes\eps)v = 1$.   
\label{prop.gauge.trivial} 
\end{thm} 
\proof 
We prove the theorem in three steps. 
 
1. $Aut(P)\cong\CA(P)$. This isomorphism has already been proven in 
Corollary~\ref{cor.auto.gauge}. We remark only that the  
map $\tilde{\theta}\sb A : Aut\sb B(P)\to \CA(P)$, defined in the 
proof of Proposition~\ref{prop.auto} has the following simple form 
$\tilde{\theta}\sb A : \CF\mapsto \Phi\sp{-1}* (\CF\circ\Phi)$. 
 
2. $\CA(P)\cong\CA(B)$. For any $\gamma\in\CA(B)$ we consider a map 
$\theta\sb B(\gamma):A\to P$ given by  
$$ 
\theta\sb B(\gamma) = \Phi\sp{-1}*\gamma*\Phi . 
$$ 
It is clear that the map $\theta\sb B(\gamma)$ is convolution 
invertible and is such that $\theta\sb B(\gamma)(1) =1$, and also  
$$ 
\Delta\sb R\theta\sb B(\gamma) = (\theta\sb B(\gamma)\otimes\id)\ad , 
$$ 
hence $\theta\sb B(\gamma)\in \CA(P)$. Now we show that the map 
$\theta\sb B:\CA(B)\to\CA(P)$ is a group homomorphism. Take any 
$\gamma\sb 1 ,\gamma\sb 2\in \CA(B)$, then   
$$ 
\theta\sb B(\gamma\sb 1*\gamma\sb 2)  =  \Phi\sp{-1}*\gamma\sb 
1*\gamma\sb 2*\Phi 
= \Phi\sp{-1}*\gamma\sb 1*\Phi*\Phi\sp{-1}*\gamma\sb 2*\Phi  
= \theta\sb B(\gamma\sb 1)*\theta\sb B(\gamma\sb 2). 
$$ 
To complete the proof of the isomorphism it remains to construct an 
inverse of $\theta\sb B$. Consider a map $\tilde{\theta}\sb B: 
\CA(P)\to Lin(A,P)$ given by   
$$ 
\tilde{\theta}\sb B(f) =\Phi*f*\Phi\sp{-1}, 
$$ 
for any $f\in\CA(P)$. It is clear that for each $f\in\CA(P)$, 
$\tilde{\theta}\sb B(f)$ is a convolution invertible map such that 
$\tilde{\theta}\sb B(f)(1)=1$. Moreover,   
\begin{eqnarray*} 
\Delta\sb R \tilde{\theta}\sb B(f)(a) & = & 
  \Phi(a\o)f(a\th)\bo\Phi\sp{-1}(a\sb{(5)})\otimes a\t f(a\th)\bt\S 
a\sb{(4)} \\   
& = &  \Phi(a\o)f(a\sb{(4)})\Phi\sp{-1}(a\sb{(7)})\otimes a\t (\S 
a\sb{(3)}) a\sb{(5)}\S a\sb{(6)}  
 =  \tilde{\theta}\sb B(f)(a) \otimes 1 . 
\end{eqnarray*} 
This proves that $\tilde{\theta}\sb B(f)\in \CA(B)$. It is now 
immediate that $\tilde{\theta}\sb B$ is an inverse of $\theta\sb B$. 
Explicitly,   
$$ 
(\tilde{\theta}\sb B\circ\theta\sb B)(\gamma) = \tilde{\theta}\sb 
B(\Phi\sp{-1}*\gamma*\Phi) = \gamma , 
$$ 
and 
$$ 
(\theta\sb B\circ\tilde{\theta}\sb B)(f) = {\theta\sb 
B}(\Phi*f*\Phi\sp{-1}) = f.  
$$ 
Therefore the required isomorphism holds. 
 
3. Finally we prove that if both $A$ and $B$ are finite dimensional 
algebras, then $\CA(B)\cong \CR(B)$. We consider a map $\theta\sb R: 
\CR(B)\to Lin(A,B)$,   
$$ 
\theta\sb R:v =\sum\sb{i\in I}v\sp i \otimes v\sb i\mapsto \gamma\sb v
=\sum\sb{i\in I} <v\sb  
i,\;\cdot\; >v\sp i, 
$$ 
where $<\; ,\; >: A\sp*\otimes A\to k$ denotes the natural pairing. 
It is clear that for each $v\in \CR(B)$, $\gamma\sb v (1) =1$, since  
$$ 
\gamma\sb v(1) = \sum\sb{i\in I}<v\sb i,1>v\sp i = (\eps\otimes\id)v =1. 
$$ 
Also, $\gamma\sb v$ is a convolution invertible map with the 
convolution inverse $\gamma\sp{-1}\sb v = \gamma\sb{v\sp{-1}}$. 
Explicitly we have   
\begin{eqnarray*} 
\gamma\sb v*\gamma\sb{v\sp{-1}}(a) & = & \sum\sb{i,j\in I} <v\sb
i,a\o><\tilde{v}\sb  
j,a\t> v\sp i\tilde{v}\sp j   
= \sum\sb{i,j\in I}<v\sb i\tilde{v}\sb j,a> v\sp i\tilde{v}\sp j \\ 
& = & <1 ,a> = \eps(a), 
\end{eqnarray*} 
where $v\sp{-1} =\sum\sb{j\in I}\tilde{v}\sp j\otimes \tilde{v}\sb j$. 
Similarly one  
shows that $\gamma\sb{v\sp{-1}}*\gamma\sb v =\eps$. Therefore for each 
$v\in\CR(B)$, we have that $\gamma\sb v\in \CA(B)$. Now we need to 
prove that $\theta\sb R$  
is a group homomorphism. For any $v,w\in \CR(B)$ we compute  
$$ 
 \gamma\sb v(a\o)\gamma\sb w(a\t) = \sum\sb{i,j\in I} \langle v\sb 
i,a\o\rangle v\sp   
i\langle w\sb j,a\t\rangle w\sp j =  
 \sum\sb{i,j\in I}\langle v\sb i w\sb j,a\rangle v\sp i w\sp j 
=  \gamma\sb{vw} (a).  
$$ 
 
For an inverse of $\theta\sb R$ we take a map $\tilde{\theta}\sb R : 
\CA(B)\to\CR(B)$ defined as follows. Let $\{ b\sp\beta; \beta\in {\cal 
B}\}$ be a basis  
of $B$. Then for any $\gamma\in \CA(B)$ and any $a\in A$ we can write   
$$ 
\gamma(a) = \sum\sb{\beta\in {\cal B}}\gamma\sb\beta(a)b\sp\beta, 
$$ 
where $\gamma\sb\beta(a)\in k$ are uniquely determined. Each 
$\gamma\sb\beta$ may be regarded as an element of $A\sp*$ such that 
$\gamma\sb\beta(a) = \langle \gamma\sb\beta , a\rangle$, for any $a\in 
A$. Since  
$\gamma\sb\beta$ are uniquely determined we can define a map   
$$ 
\tilde{\theta}\sb R: \gamma\mapsto\sum\sb{\beta\in {\cal B}} 
b\sp\beta\otimes \gamma\sb\beta .  
$$ 
These are the elementary facts that if $\gamma\in \CA(B)$, then 
$\tilde{\theta}\sb R(\gamma)\in\CR(B)$ and that $\tilde{\theta}\sb R$ 
is an inverse of $\theta\sb R$.   
 
This completes the proof of the theorem. \endproof 
 
Therefore 
Theorem~\ref{prop.gauge.trivial} allows one to interpret a vertical 
automorphism of a (locally) trivial quantum principal bundle as 
a change of local variables and truly as a gauge transformation of a 
trivial quantum principal bundle. Furthermore it gives a geometric 
interpretation of a universal $R$-matrix of a quasitriangular Hopf 
algebra $H$ \cite{drinfeld1} as a gauge transformation of the quantum 
frame bundle  
$D(H)\sp*(H,A)$ \cite[Example~5.6]{brzezinski6}, with a total space dual to 
Drinfeld's double $D(H)$ and the structure quantum group $A=H\sp*$  
(cf.~\cite[Section~5.7.2]{brzezinski14}). 
 
Gauge transformations have also a natural interpretation in terms 
of isomorphisms of crossed product algebras. Recall from  
\cite{brzezinski6,brzezinski14} that a total space 
$P$ of a trivial quantum principal bundle $P(B,A,\Phi)$ is isomorphic 
to $B\otimes A$ as a vector space. Define the crossed product 
algebra $B\sb\Phi\rtimes A$ by equiping $B\otimes A$ with a multiplication 
$$ 
(b\sb 1\otimes a\sp 1)(b\sb 2\otimes a\sp 1) =   
b\sb 1 \Phi(a\sp 1\o) b\sb 2 \Phi(a\sp 2\o)\Phi\sp{-1}(a\sp 1\t a\sp 2\t) 
\otimes a\sp 1\th a\sp 2\th. 
$$ 
Then $P\cong B\sb\Phi\rtimes A$ as algebras. The isomorphism is given 
by $\theta\sb\Phi: u\mapsto u\bo\Phi\sp{-1}(u\bt\o)\otimes u\bt\t$. The 
following proposition is a special case of the result of Doi \cite{doi1} 
(see also \cite[Proposition 4.2]{majid1}).

\begin{prop} 
Let $P(B,A,\Phi)$ be a trivial quantum principal bundle. Let for any 
trvialisation $\Psi$ of $P(B,A,\Phi)$, $\Theta\sb\Psi : 
B\sb\Psi\rtimes A\to B\sb\Phi\rtimes A$ be a crossed product algebra 
isomorphism such that $\Theta\sb\Psi\mid\sb B = \id$ and $\Delta\sb R\Theta 
\sb\Psi = (\Theta\sb\Psi\otimes\id)\Delta\sb R$. Then there is a  
bijective correspondence between all isomorphisms $\Theta\sb\Psi$ 
corresponding to all trivialisations $\Psi$ and the gauge transformations 
of $P(B,A,\Phi)$. 
\label{prop.also.shahn} 
\end{prop} 
 
\section{An Example} 
In this section we consider the simplest example of a trivial quantum 
principal bundle for which the gauge group can be computed explicitly. 
This example serves as an illustration of the considerations of 
the previous section and in particular of the use of  
Theorem~\ref{prop.gauge.trivial} in computations of the gauge group. 
It also shows that the quantum gauge group ${\cal A}(B)$ of a 
classical trivial principal bundle is much bigger than the 
classical gauge group of this bundle. We assume that $k = \bf C$. 
 
\begin{prop} 
Let $B$ be an $M$-dimensional semisimple algebra with unit and let 
$A={\bf C}[G]$ be a group algebra of a finite group $G$ of $N$ elements.  
Let $\rho\sb l$ , $l=0, \ldots, L-1$ be all   
non-equivalent irreducible representations of $B$.  
Then for any trivial quantum principal bundle 
$P(B,A,\Phi)$,  
$$ 
\CA(B)\cong \underbrace{\CG\oplus \CG\oplus \ldots \oplus \CG}\sb{N-1}, 
$$ 
where  
\begin{equation} 
\CG = \{ b\in B; b \; {\rm is\;  invertible\; in\;} B\} = \{ b\in B;
\prod\sb{l=0}\sp {L-1} \det\rho\sb l(b)\neq 0\}.  
\label{group} 
\end{equation} 
\label{prop.gauge.finite.group} 
\end{prop} 
\proof Since the gauge group depends only on $A$ and $B$ it suffices 
to consider the case $P=B\otimes A$. Let $b\sb 0=1, b\sb 1\ldots , b\sb{M-1}$ 
be a basis of $B$, and $g\sb 0 =1, g\sb 1,\ldots , 
g\sb{N-1}$ be elements of $G$. Assume that the multiplication 
in $B$ is given by 
\begin{equation} 
b\sb i b\sb j = \sum\sb{k=0} N\sb{ij}\sp k b\sb k, \qquad N\sb{ij}\sp k  
\in\C . 
\label{prod.b} 
\end{equation} 
First we show that the gauge group $\CA(B)$ is a direct sum of identical 
groups. Let $\gamma\in\CA(B)$ be given by $\gamma(g\sb k ) = \sum\sb{i=0} 
\sp{M-1} a\sb{k}\sp{i}b\sb i$, for some $a\sb{k}\sp i\in\C$. By  
Theorem~\ref{prop.gauge.trivial} there is a unique vertical  
automorphism $\CF :B\otimes A\to B\otimes A$, canonically  
associated to $\gamma$. Since all 
generators of $A$ are group-like $\CF$ can be easily computed,  
$$ 
\CF(b\sb i\otimes g\sb k) = \sum\sb{j,l=0}\sp{M-1} a\sp j\sb k 
N\sb{ij}\sp lb\sb l 
\otimes g\sb k. 
$$ 
Since $\CF$ does not act on elements of $G$, for each $k = 0, \ldots ,N-1$, 
it may be 
regarded as a linear automorphism of $B$. Noting that for $k=0$, 
$\CF$ is the identity map we thus obtain 
$$ 
\CA(B)\cong Aut\sb B(B\otimes A) =  
\underbrace{\CG\oplus \CG\oplus \ldots \oplus \CG}\sb{N-1},  
$$ 
where $\CG$ may be identified with a group of all  non-singular 
$M\times M$ matrices $F({\bf a}) = (F({\bf a})\sb i\sp j)\sb{i,j =0}\sp{M-1}$ 
with the entries 
$$ 
F({\bf a})\sb i\sp j = \sum\sb{k=0}\sp{M-1} a\sp kN\sb{ik}\sp j . 
$$ 
Next we observe that the map $F: B\to End(B)$, $F: \sum\sb{i=0}\sp{M-1} 
a\sp ib\sb i \mapsto F({\bf a})$ is a right regular representation of
$B$.  Since  
$B$ has a unit $F$ is an algebra isomorphism. Therefore 
the matrix $F({\bf a})$ is non-singular if and only if  $\sum\sb{i=0}\sp{M-1} 
a\sp ib\sb i$ is invertible. This gives the first description of
$\CG$. Secondly we  
observe that because $B$ is semisimple $F$ is completely reducible and it 
contains every irreducible representation of $B$ at least once. Therefore $ 
\det F({\bf a}) \neq 0$ iff $ \prod\sb{l=0}\sp{L-1} 
\det\rho\sb l(\sum\sb{i=0}\sp{M-1} a\sp ib\sb i) \neq 0$. This establishes 
the second description of $\CG$.  
\endproof 
\begin{ex} 
\rm 
Let $P = \C[{\bf Z}\sb{MN}] \equiv \C[1, g]/(g\sp{MN} -1)$ and 
$A= \C[{\bf Z}\sb{N} ]\equiv \C[1, h]/(h\sp N -1)$ have the standard 
Hopf algebra structure, i.e. $\Delta g = g\otimes g$, $\Delta h =  
h\otimes h$ etc. Define a Hopf algebra projection $\pi: P\to A$, 
$\pi: g\mapsto h$. Let $\Delta\sb R : P\to P\otimes A$ be a right 
coaction given by a pushout $\Delta\sb R = (\id\otimes\pi)\circ\Delta$. 
Then $B = P\sp A$ is a subalgebra of $P$ as a vector space spanned 
by $\{1, g\sp N,\ldots , g\sp{N(M-1)}\}$ and hence is isomorphic to 
$\C[{\bf Z}\sb{M}]$. $B$ is a semisimple algebra with unit as a group 
algebra. Clearly $P(B,A,\pi)$ is a trivial quantum principal 
bundle on the homgeneous space $B$ with trivialisation given by 
$\Phi\sb 0: A\to P$, $\Phi\sb 0: h\sp n\mapsto g\sp n$, $0\leq n\leq N-1$. 
 
Since $B$ is a commutative algebra all the irreducible   
representations  $\rho\sb k$, $k=0,\ldots , M-1$  
 of $B$  are one dimensional, 
 \begin{equation} 
\rho\sb k (g\sp{mN}) = e\sp{{2\pi km\over M}i}. 
\label{rho.rep} 
\end{equation} 
Therefore $b = a\sb m g\sp{mN}$ is an element of $\CG$ iff 
$\forall k\in\{ 0,\ldots , M-1\}$, $\sum\sb{m=0}\sp{M-1} a\sb m  
e\sp{{2\pi km\over M}i}\neq 0$. 
 
To gain the further inside into the structure of $\CG$ we consider 
the group homomorphism $\rho: \CG\to (\C\sp*)\sp M = \C\sp*\oplus\ldots\oplus 
\C\sp*$, $\rho: 
b\mapsto (\rho\sb 0(b),\ldots, \rho\sb{M-1}(b))$. Let $b= 
\sum\sb{m= 0}\sp{M-1}a\sb mg\sp{mN}$ and $c\sb k =  
\sum\sb{m=0}\sp{M-1} a\sb m  
e\sp{{2\pi km\over M}i} = \rho\sb k(b)$. Clearly the map $\rho$ 
is injective. It is also surjective since the Vandermonde 
determinant $\det(e\sp{{2\pi km\over M}i})\sb{k,m =0}\sp{M-1}$ is 
non-zero. Therefore $\rho$ is an isomorphism of multiplicative 
groups, $\CG \cong (\C\sp*)\sp M$, and $\CA(B) = (\C\sp*)\sp{M(N-1)}$.  
 
Using the properties of the Vandermonde determinants 
\cite[pp. 322 and 333]{muir1} one easily finds that the 
inverse of $\rho$, $\rho\sp{-1}: (c\sb 0, \ldots ,c\sb{M-1}) \mapsto 
\sum\sb{m=0}\sp{M-1}a\sb m g\sp{mN}$ is given by 
\begin{equation} 
a\sb m \!\! = \!\!(-1)\sp m \sum\sb{k=0}\sp{M-1} 
\prod\sb{\stackrel{j = 0}{j\neq k}}\sp{M-1}\left(e\sp{2\pi {j\over M}i} 
-e\sp{2\pi  {k\over M}i}\right)\sp{-1}  
\sum\sb{\bf m} 
\delta\sb{0 m\sb{k}}c\sb k\exp\left( 2\pi i {1\over M}  
\sum\sb{l=1}\sp{M-1} lm\sb l\right), 
\label{solution} 
\end{equation} 
where the second sum runs over all sequences ${\bf m} =  
(m\sb 0, \ldots ,m\sb{M-1})$, 
$m\sb l =0,1$,  such that $m\sb 0 +\ldots +m\sb{M-1} = M-1-m$.  
\label{example.cyclic} 
 
The quantum principal bundle discussed in this example may 
serve as an illustration of the fact that the quantum gauge group 
$\CA(B)$ of a classical principal bundle is much bigger that 
the classical gauge group $\CA\sb{alg}(B)$, whose elements are 
algebra maps $A\to B$. 
 
To compute $\CA\sb{alg}(B)$ we notice first that a gauge transformation  
$\gamma\in {\cal A}\sb{alg}(B)$ is fully  
determined by its action on $h$, hence $\CA(B)\sb{alg}$ may be 
identified with a subgroup of $\CG = (\C\sp*)\sp M$.  
Secondly $\gamma(h)\sp N =1$ because $\gamma$ is an algebra 
map. When $\gamma$ is viewed as $(c\sb 0, \ldots, c\sb{M-1}) 
\in (\C\sp*)\sp M$ via the isomorphism $\rho$, the 
condition $\gamma(h)\sp N =1$ is equivalent to $c\sb k\sp N =1$. 
Hence each $c\sb k$ is an element of $\Z\sb N$ viewed 
as a multiplicative subgroup of $\C\sp*$ by $\Z\sb N\ni n\mapsto 
e\sp{2\pi n i/N}\in \C\sp*$. Therefore 
$$ 
{\cal A}\sb{alg}(B)\cong\underbrace{{\bf Z}\sb N\oplus  
{\bf Z}\sb N \oplus\ldots 
\oplus {\bf Z}\sb N}\sb{M}. 
$$ 
 
Finally we can show that the bundle $P(B,A,\pi)$  
of Example~\ref{example.cyclic} is a classical 
trivial bundle in the sense that it admits a trivialisation which is an  
algebra map. Classifying all such trivialisations we classify  
all cross-sections of $P(B,A,\pi)$ 
which are algebra maps by Proposition~\ref{prop.section.trivial}. 
 
The algebraic trivialisation $\Phi$ is fully determined by 
its action on $h$ and also it must be related to $\Phi\sb 0$ by 
a gauge transformation. Therefore we must have  
$\Phi(h) = \sum\sb{m=0}\sp{M-1} a\sb{m}g\sp{mN +1}$, for some 
$a\sb{m}\in \C$. The trivialisation $\Phi$ 
is an algebra map if and only if $\Phi(h)\sp N =1$.  
This condition may be easily solved if we first notice 
that it must be satisfied in all representations $\rho\sb k$ and 
then linearise obtained system of equations by taking its 
$N$-th root. As a result we obtain the $N\sp M$ systems of $M$ linear  
equations parametrised by the sequences $(n\sb 0, \ldots ,n\sb{M-1})$, 
$n\sb k =0, \ldots, N-1$. Each such system may be easily solved 
using the properties of the Vandermonde determinants. We obtain 
$$ 
a\sb{m} \!\! = \!\!(-1)\sp m \sum\sb{k=0}\sp{M-1} 
\prod\sb{\stackrel{j = 0}{j\neq k}}\sp{M-1}\left(e\sp{2\pi i{j\over M}} 
-e\sp{2\pi i {k\over M}}\right)\sp{-1}  
\sum\sb{\bf m} 
\delta\sb{0 m\sb{k}}\exp\left( 2\pi i\left( {1\over M}  
\sum\sb{l=1}\sp{M-1} lm\sb l + {n\sb k\over N} - {k\over{MN}}\right)\right), 
$$ 
where the range of the second sum is over all sequences ${\bf m} = (m\sb 0, 
\ldots , m\sb{M-1})$ as in (\ref{solution}).  
 
We notice that for some choices of $M$ and $N$, for example $M=N=2$, there 
are no real solutions $a\sb{m}$ to the equation $\Phi(h)\sp N =1$. 
Therefore in these cases the bundle $P(B,A,\pi)$ can be considered as a  
classical trivial bundle over $\bf C$ and as a trivial quantum 
bundle and nontrivial classical bundle over $\bf R$. 
\dia 
\end{ex} 
\begin{rk} 
\rm Interesting examples of commutative semisimple algebras $B$ come
from the fusion rings of  
algebraic quantum field theories. A fusion ring is a ring with
involution generated by $b\sb i$  
subject to the relation (\ref{prod.b}) with $N\sb{ij}\sp k\in {\bf
Z}$.  To each $b\sb i$ one can  
associate a non-negative number $d(b\sb i)$ which has a meaning of a quantum 
dimension of a representation of an internal braided (or
quasi-quantum)  symmetry group, corresponding  
to $b\sb i$. By the theorem of Rehren \cite[Proposition~3.3]{rehren1}
an element $b\sb i$  
of a fusion ring is invertible iff $d(b\sb i)=1$. We can extend fusion
ring to an algebra  
$B$ over $\C$ and we can extend linearly the function $d$ to
$B$. Therefore for such an  
algebra $B$,  all elements of $\CG$ have non-zero quantum dimension. 
The representation theory of fusion algebras is well-understood 
in many cases, hence in all those cases $\CG$ can be explicitly 
determined. 
 
The algebra $B$ of Example~\ref{example.cyclic} is a fusion algebra of
rational gaussian  
models of conformal field theory \cite{verlinde1}.\footnote{I would
like to thank Matthias Gaberdiel for a discussion.} \dia  
\end{rk}

\vglue 0.4cm 
{\elevenbf \noindent  Acknowledgements \hfil} 
\vglue 0.3cm 
\noindent I am grateful to  Piotr Hajac and Shahn Majid for discussions  
and comments. This work was supported by the EU HCM grant and the grant 
KBN 2 P 302 21706 p 01.  
 
\end{document}